\documentstyle[11pt,paspconf,epsf]{article}
\begin{document}

\title{The Halo Formation Rate and its link to the Global Star Formation Rate}
\author{Will Percival, Lance Miller, Bill Ballinger}
\affil{Astrophysics, Department of Physics, Keble Road, Oxford, OX1 3RH}

\begin{abstract}
The star formation history of the universe shows strong evolution with
cosmological epoch. Although we know mergers between galaxies can
cause luminous bursts of star formation, the relative importance of
such mergers to the global star formation rate (SFR) is unknown. We
present a simple analytic formula for the rate at which halos merge to
form higher-mass systems, derived from Press-Schechter theory and
confirmed by numerical simulations (for high halo masses). A
comparison of the evolution in halo formation rate with the observed
evolution in the global SFR indicates that the latter is largely
driven by halo mergers at $z>1$. Recent numerical simulations by
Kolatt et al.\ (1999) and Knebe \& M\"{u}ller (1999) show how merging
systems are strongly biased tracers of mass fluctuations, thereby
explaining the strong clustering observed for Lyman-break galaxies
without any need to assume that Lyman-break galaxies are associated
only with the most massive systems at $z\sim3$.
\end{abstract}

\section{Calculating the Halo Formation Rate}

In our analysis, a halo formation event is considered to have occurred
when all the mass in the halo is assembled. Note that this is
different to that previously used by some authors (Lacey \& Cole
1993). We wish to answer the question: `Given a halo of mass M forms
at some time, what is the probability $P(t|M)$dt that it forms in the
time interval (t,t+dt)?'
\begin{figure}
  \plotone{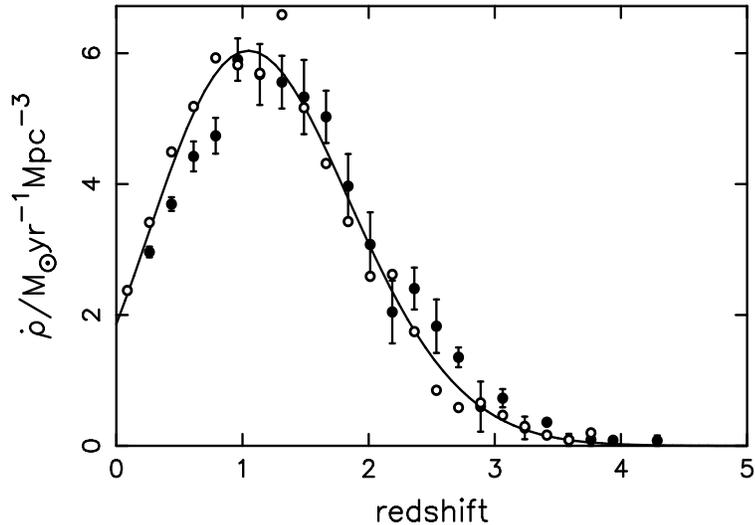} 
  \caption{Comparison of N-body results (solid circles) with
  Press-Schechter predictions of the halo formation rate for halos of
  mass $\sim1.3\times10^{13}{\rm\,M_\odot}$. The curve shows the
  prediction of equation~\protect\ref{eq:rate_cstm} at this mass with
  parameters as given in the text and normalised to the N-body
  values. The distribution of $10^4$ `formation events' at the
  required mass from Monte-Carlo realisations of random walks is also
  shown (open circles).}
\label{fig:nbody}
\end{figure}

Standard PS theory calculates $P(M|t)$dM, the distribution of halo
mass at fixed epoch, and we have shown that it is possible to
calculate $P(t|M)$dt from this using Bayes' theorem. We can also
calculate the same formula using intrinsic properties of Brownian
random walks invoked in PS theory. The mass of halo a small volume
element resides in at time t, is given by the first upcrossing of the
line $\delta=\delta_c$ by a Brownian random walk in
($\delta$,$\sigma_M^2$) space, where $\delta$ is a function of time
and $\sigma_M^2$ is a function of mass. Using the theory of random
walks we can calculate the distribution of first upcrossing times at
$\sigma_M^2$, P($\delta_c|\sigma_M^2$), from which a simple change of
variables can be used to obtain $P(t|M)$dt:
\begin{equation}
  P(t|M)dt=\frac{\delta_c}{\sigma_M^2}
    \exp\left(-\frac{\delta_c^{2}}{2\sigma_M^2}\right)
    \left|\frac{d\delta_c}{dt}\right|dt. \label{eq:rate_cstm}
\end{equation}

This equation is in good agreement with Monte-Carlo realisations of
Brownian random walks (Fig.~\ref{fig:nbody}). We have also run a large
N-body simulation, using the Hydra N-body hydrodynamics code
(Couchman, Thomas \& Pearce 1995). Groups of between 45 and 47
particles ($1.3\times10^{13}{\rm\,M_\odot}$) were identified using a
standard friends-of-friends algorithm at 362 output times. The number
which could have formed in each time interval is compared to the
expected distribution in Fig.~\ref{fig:nbody}.

Now suppose we are only interested in a subset of formation events -
e.g. those which involve similar mass objects merging together. The
formation rate from such mergers is the same as that derived above,
because all walks which pass through a given point can be thought of
as new walks starting from that point. Consequently the mass
distribution of progenitors immediately prior to the formation event
is independent of the formation epoch. Turning this argument around,
placing constraints on the progenitors of halos immediately prior to
their formation doesn't affect the distribution of formation times,
although this might affect the bias (see later).

\section{The Star Formation Rate}

It is interesting to compare the halo formation rate with the observed
global star formation rate.  At $z<1$ the rate of halo formation falls
off very rapidly with cosmic time, independently of halo mass, and it
is likely that this effect is primarily responsible for the observed
rapid evolution in the CFRS.  Such strong evolution is not seen in
semi-analytic models that only include a quiescent component of star
formation (Guiderdoni et al. 1998).

\begin{figure}
  \plotone{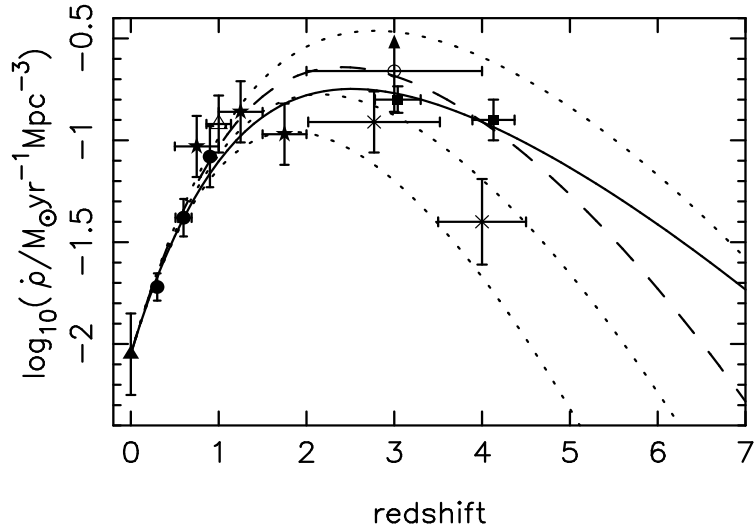} 
  \caption{The observed mean comoving volume-averaged SFR as
  determined from the CFRS (Lilly et al.\ 1996) (solid circles),
  optical HDF data (Madau et al.\ 1996; Pettini et al.\ 1997)
  (crosses) and Connolly et al. (1997) (solid stars),
  extinction-corrected Lyman break galaxies (Steidel et al.\ 1998)
  (solid squares), sub-mm data (Hughes et al.\ 1998) (open circle),
  and H$\alpha$ surveys (Gallego et al.\ 1995) (solid triangle) and
  (Glazebrook et al.\ 1998) (open triangle). We assume a Salpeter IMF
  and flat $\Omega_{M}=1$ cosmology. Dotted lines are the predicted
  halo formation rates normalised to the local SFR (convolved with a
  nominal starburst lifetime of 0.6\,Gyr, Bruzual et al. 1993) for
  masses of $10^{10.0}, 10^{11.0}, 10^{11.5} {\rm\,M_\odot}$ and for a
  mass of $10^{10.6} {\rm\,M_\odot}$ (dashed line). As an
  illustration, we combine the formation rate of different mass halos
  weighted by a Gaussian in dn/dlogM centred at a mass of
  $10^{10.6}{\rm\,M_\odot}$ (solid line). The low-z evolution is
  unaffected, but an increasing contribution from lower-mass halos
  flattens the curve at high z.}
\label{fig:sfr}
\end{figure}

\section{The Clustering of Lyman-break Galaxies}

The strong clustering of Lyman-break galaxies has been explained as
being due to the high bias of the most massive overdensities, and
consideration of the abundance of Lyman-break galaxies and their
clustering leads to an interpretation of them as being associated with
massive halos, $M \sim 8 \times 10^{11} h^{-1} {\rm\,M_\odot}$
(Adelberger et al. 1998) for an $\Omega=0.3$ flat universe. If star
formation is initiated after halo formation then these most massive
halos form too late in the universe to reproduce the observed
evolution in SFR in a simple way: we would require the efficiency of
star-formation to evolve with redshift. Conversely, if the Lyman-break
galaxies have lower mass but are associated with newly-formed, merging
halos, then we might suppose that such mergers are also highly biased,
and recent simulations indicate this to be the case (Kolatt et
al. 1999, Knebe \& M\"{u}ller 1999).

\section{Conclusions}

We have presented the key points involved in deriving a simple formula
for the rate of formation of new halos using Press-Schechter
theory. It agrees with Monte-Carlo and N-body simulation results. We
have argued that the strong cosmological evolution observed in the SFR
is primarily driven by the cosmic variation in the rate of halo
formation.  Given that quiescent star formation does not provide
enough evolution (Guiderdoni et al. 1998) we suggest that
merger-induced starbursts are extremely important for star formation
at $z \sim 1$ and are perhaps the principal sites of the observed star
formation at high redshifts. At high z, a more physically-motivated
model is needed to deduce the relative contributions of a range of
halo masses, but we have shown that a simple combination of such a
range can produce evolution consistent with present data. Recent
results indicate that such merging-halo systems are also sufficiently
highly biased to explain the strong clustering of Lyman-break galaxies
at $z \sim 3$.

The work highlighted here is more comprehensively covered in our
recent paper available as astro-ph/9906204. We are also continuing to
work on the bias of merging halos from numerical simulations.

\end{document}